\RequirePackage[mathlines]{lineno} 
\documentclass[a4paper,11pt]{article}

\usepackage{jheppub} 
\usepackage[T1]{fontenc} 
\emailAdd{besiii-publications@ihep.ac.cn}
\usepackage{threeparttable}
\usepackage{multirow}
\usepackage{subfigure}
\usepackage{float}
\let\oldequation\equation
\let\oldendequation\endequation
\renewenvironment{equation}
  {\linenomathNonumbers\oldequation}
  {\oldendequation\endlinenomath}

\def \ee   {e^+e^-}

\def \mev  {\mbox{MeV}}

\begin{document}

\title{\boldmath Measurement of branching fractions of $\Lambda_{c}^{+}$ decays to $\Sigma^{+} K^{+} K^{-}$, $\Sigma^{+}\phi$ and $\Sigma^{+} K^{+} \pi^{-}(\pi^{0})$}

\collaboration{The BESIII collaboration}

\abstract{
 Based on 4.5 fb$^{-1}$ data taken at seven center-of-mass energies ranging from 4.600 to 4.699 GeV with the BESIII detector at the BEPCII collider, we measure the branching fractions of $\Lambda_{c}^{+}\rightarrow\Sigma^{+}+hadrons$ relative to $\Lambda_{c}^{+}\rightarrow \Sigma^+ \pi^+ \pi^-$. Combining with the world average branching fraction  of $\Lambda_{c}^{+}\rightarrow \Sigma^+ \pi^+ \pi^-$, their branching fractions are measured to be $(0.377\pm0.042\pm0.020\pm0.021)\%$ for $\Lambda_{c}^{+}\rightarrow\Sigma^{+} K^{+} K^{-}$, $(0.200\pm0.023\pm0.011\pm0.011)\%$ for $\Lambda_{c}^{+}\rightarrow\Sigma^{+} K^{+} \pi^{-}$, $(0.414\pm0.080\pm0.030\pm0.023)\%$ for $\Lambda_{c}^{+}\rightarrow\Sigma^{+}\phi$ and $(0.197\pm0.036\pm0.009\pm0.011)\%$ for $\Lambda_{c}^{+}\rightarrow\Sigma^{+}K^{+} K^{-}$(non-$\phi$). In all the above results, the first uncertainties are statistical, the second are systematic and the third are from external input of the branching fraction of $\Lambda_{c}^{+}\rightarrow \Sigma^+ \pi^+ \pi^-$. Since no signal for $\Lambda_{c}^{+}\rightarrow\Sigma^{+} K^{+} \pi^{-}\pi^{0}$ is observed, the upper limit of its branching fraction is determined to be 0.13\% at the 90$\%$ confidence level. 
}


\arxivnumber{XXXXXXX}

\maketitle
\flushbottom

\section{INTRODUCTION}
\label{sec:introduction}
\begin{figure}[!hbt]
\centering
	\subfigure[]
 { 
	\includegraphics[width=0.27\paperwidth]{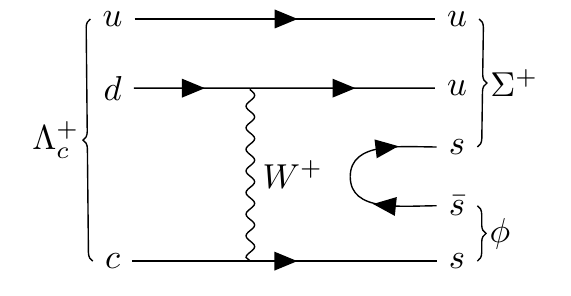}
	\label{fig:fphi}
		 } 
	\hspace{0.5pt}
	\subfigure[]
  { 
	\includegraphics[width=0.24\paperwidth]{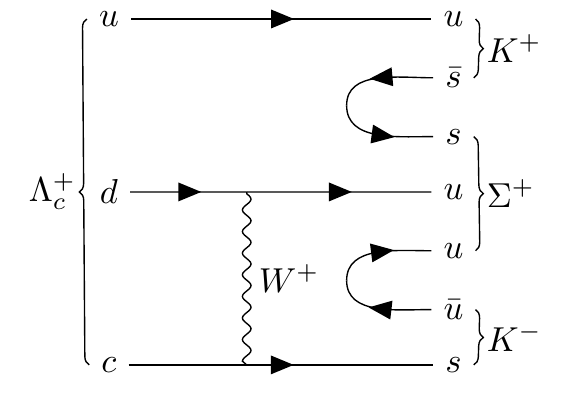}
	\label{fig:fkk}
		 } 
	\hspace{0.5pt}
\subfigure[]
{ 
	\includegraphics[width=0.24\paperwidth]{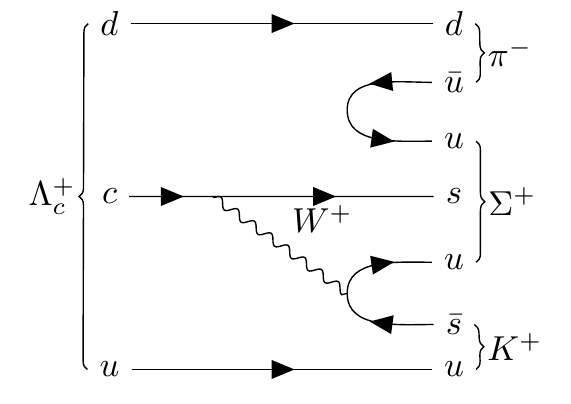}
	\label{fig:fkpi}
	 } 
	\hspace{0.5pt}
	\subfigure[]

{ 
	\includegraphics[width=0.24\paperwidth]{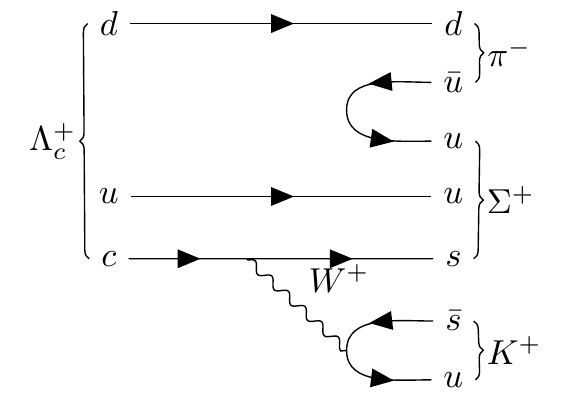}
	\label{fig:fkpi2}
		 } 
\caption{Feynman diagrams of (a)$\Lambda_{c}^{+}\rightarrow\Sigma^{+}\phi$, (b)$\Lambda_{c}^{+}\rightarrow\Sigma^{+} K^{+} K^{-}$ and (c)(d)$\Lambda_{c}^{+}\rightarrow\Sigma^{+} K^{+} \pi^{-}$ processes.}
\label{fig:feydia}
\end{figure}

\hspace{1.5em} 
Weak decays of charmed baryons provide useful information for understanding the interplay of weak and strong interactions, complementary to the information obtained from charmed mesons~\cite{Klein:1989tj,Asner:2008nq}. The lightest charmed baryon $\Lambda_c^+$ is the cornerstone of all charmed baryon spectroscopy, and the measurement of the properties of the $\Lambda_c^+$ provides essential inputs for studying other heavy-quark baryons, such as the singly- and doubly-charmed baryons~\cite{Yu:2017zst} and $b$-baryons~\cite{Rosner:2012gj}. Contrary to the significant progress made in the studies of the heavy meson decay, advancement in the arena of heavy baryons, on both theoretical and experimental sides, has been very slow~\cite{Cheng:2015iom}. 
Since 2014, much progress has been made by BESIII, Belle and LHCb experiments in studying $\Lambda_c^+$ decays. This includes measurements of absolute branching fractions (BFs) and observation of new $\Lambda_c^+$ decay modes ~\cite{BESIII:lmdenv, BESIII:hadron, BESIII:phh, BESIII:lmdmunv,BESIII:sigmapipi, BESIII:nkspi, BESIII:LmdX, BESIII:eX, BESIII:xik, BESIII:Lmdpieta,  BESIII:KsX,  BESIII:pkseta, BESIII:npi, Belle:pkpi, Belle:pkpi_dcs, Belle:sigmapipi, Belle:ppi0peta, Belle:pomega,Li:2021iwf}, measurements of weak decay  asymmetries of $\Lambda_c^+$ decays ~\cite{BESIII:2019odb,Belle:2022uod}, determination of $\Lambda_c^+$ spin~\cite{BESIII:2020kap}, lifetime~\cite{Belle-II:2022ggx,LHCb:2019ldj} and search for the rare $\Lambda_c^+$ decays~\cite{LHCb:pomega}.  All of this supplies rich data for improving theoretical models~\cite{Cheng:2021qpd}.   
The external $W$-emission process is factorizable, but internal $W$-emission 
and $W$-exchange processes are not, and non-factorizable contributions 
have sizeable theoretical uncertainties.  
The $W$-exchange process is particularly relevant for charmed baryons, 
where it is not suppressed by helicity and color as in the meson case.  
Therefore, more precise measurements of the $W$-exchange decays 
of the $\Lambda_c^+$ play an important role 
in the identification of the non-factorizable contribution 
in different theoretical calculations~\cite{Cheng:2021qpd,Belle:2001hyr}.  
The two decay modes $\Lambda_{c}^{+}\rightarrow\Sigma^{+}\phi$ and $\Lambda_{c}^{+}\rightarrow\Sigma^{+} K^{+} K^{-}$ are expected to proceed entirely through non-factorizable $W$-exchange diagrams \cite{Cheng:1993gf}, 
as shown in Figures~\ref{fig:fphi} and ~\ref{fig:fkk}. 

Complementary studies of charmed baryons, such as the problem of the $\Xi_c^+$ lifetime~\cite{Guberina:2002fz}, are considered in the framework of Heavy-Quark Expansion~\cite{Khoze:1983yp}, which can be expressed in terms of measurable inclusive quantities of the other two charmed baryons belonging to the same SU(3) flavor multiplet in a model-independent way. In such a treatment, the inclusive decay rates of singly Cabibbo-suppressed decay modes play a prominent role. Two such modes, $\Lambda_{c}^{+}\rightarrow\Sigma^{+} K^{+} \pi^{-}$ and $\Lambda_{c}^{+}\rightarrow\Sigma^{+} K^{+} \pi^{-}\pi^{0}$, are expected to proceed entirely through internal and external  $W$-emission diagrams as shown in  Figures~\ref{fig:fkpi} and \ref{fig:fkpi2}, respectively. Measurements of the $\Lambda_{c}^{+}$ singly Cabibbo-suppressed decay rates can improve our understanding of the lifetime hierarchy \cite{LHCb:2018nfa,LHCb:2021vll,Cheng:2021vca}. The BFs of both $\Lambda_{c}^{+}\rightarrow\Sigma^{+} K^{+} \pi^{-}$ and $\Lambda_{c}^{+}\rightarrow\Sigma^{+} K^{+} K^{-}$ are predicted to be  $(0.025\pm 0.03)\%$~\cite{Cen:2019ims}, and $\Lambda_{c}^{+}\rightarrow\Sigma^{+}\phi$ is predicted to be $(0.39\pm 0.06)\%$~\cite{Hsiao:2019yur}. 

The ratios of the BFs (RBFs) of $\Lambda_{c}^{+}\rightarrow\Sigma^{+}\phi$ and $\Lambda_{c}^{+}\rightarrow\Sigma^{+} K^{+} K^{-}$ relative to $\Lambda_{c}^{+}\rightarrow p K^{-} \pi^{+}$ were measured by CLEO to be $0.069\pm 0.023\pm 0.016$ and $0.070\pm 0.011\pm 0.011$ respectively~\cite{CLEO:1993fhs}. In 2002, Belle reported the RBFs of $\Lambda_{c}^{+}\rightarrow\Sigma^{+} K^{+} K^{-}$, $\Lambda_{c}^{+}\rightarrow\Sigma^{+}\phi$ and $\Lambda_{c}^{+}\rightarrow\Sigma^{+}K^{+}\pi^{-}$ relative to $\Lambda_{c}^{+}\rightarrow\Sigma^{+} \pi^{+} \pi^{-}$, which are $0.076\pm 0.007\pm 0.009$,
$0.085\pm 0.012\pm 0.012$ and $0.047\pm 0.011\pm 0.008$, respectively~\cite{Belle:2001hyr}.

In this work, we present the measurements of the RBFs of the signal decays $\Lambda_{c}^{+}\rightarrow\Sigma^{+} K^{+} K^{-}$, $\Sigma^{+} K^{+} \pi^{-}$, $\Sigma^{+}\phi$, $\Sigma^{+} K^{+} K^{-}$(non-$\phi$) and $\Sigma^{+} K^{+} \pi^{-}\pi^{0}$, relative to the reference decay $\Lambda_{c}^{+} \rightarrow \Sigma^{+} \pi^+ \pi^-$, by analyzing 4.5 fb$^{-1}$ data taken at the center-of-mass energies $\sqrt{s}=4.600$, 4.612, 4.628, 4.641, 4.661, 4.682 and 4.699~GeV~\cite{BESIII:2022ulv} with the BESIII detector at the BEPCII collider. Throughout this paper, charge-conjugate modes are implicitly assumed. In Sect.~\ref{sec:detector}, the BESIII detector and the data samples used in this analysis are described. The event selection is introduced in Sect.~\ref{sec:analysis} and the determination of the BF is presented in Sect.~\ref{sec:bf}. The systematic uncertainties are discussed in Sect.~\ref{sec:systematic}. Finally, Sect.~\ref{sec:summary} summarizes the results. 

\begin{table}[!htbp]
\footnotesize
   \caption{The center-of-mass energy $E_{\rm cms}$ and the integrated luminosity measured $L$ we used in this analysis.}
   \begin{center}
   \begin{tabular}{ccc}
        \hline
        \hline
	   $Sample$  &$E_{\rm cms}$ ($\mev$)  & $L$ (pb$^{-1}$)\\
	\hline
    4.600& 4599.53$\pm$0.07$\pm$0.74&~~586.90$\pm$0.10$\pm$3.90\\
    4.612& 4611.86$\pm$0.12$\pm$0.30&~~103.65$\pm$0.05$\pm$0.55\\
	4.628& 4628.00$\pm$0.06$\pm$0.32&~~521.53$\pm$0.11$\pm$2.76\\
    4.641& 4640.91$\pm$0.06$\pm$0.38&~~551.65$\pm$0.12$\pm$2.92\\
	4.661& 4661.24$\pm$0.06$\pm$0.29&~~529.43$\pm$0.12$\pm$2.81\\
    4.682& 4681.92$\pm$0.08$\pm$0.29& ~1667.39$\pm$0.21$\pm$8.84\\
    4.699& 4698.82$\pm$0.10$\pm$0.36&~~535.54$\pm$0.12$\pm$2.84\\
	\hline 
        \hline
   \end{tabular}
    \end{center}
	 \label{tab:sum}
\end{table}
\section{BESIII DETECTOR AND MONTE CARLO SIMULATION}
\label{sec:detector}
\hspace{1.5em}
The BESIII detector~\cite{Ablikim:2009aa} records symmetric $e^+e^-$ collisions 
provided by the BEPCII storage ring~\cite{Yu:IPAC2016-TUYA01}, which operates with a peak luminosity of $1\times10^{33}$~cm$^{-2}$s$^{-1}$
in the center-of-mass energy range from 2.0 to 4.95~GeV.
BESIII has collected large data samples in this energy region~\cite{Ablikim:2019hff}. The cylindrical core of the BESIII detector covers 93\% of the full solid angle and consists of a helium-based
 multilayer drift chamber~(MDC), a time-of-flight system~(TOF), and a CsI(Tl) electromagnetic calorimeter~(EMC),
which are all enclosed in a superconducting solenoidal magnet
providing a 1.0~T magnetic field. The solenoid is supported by an
octagonal flux-return yoke with resistive plate counter muon
identification modules interleaved with the steel. 

The charged-particle momentum resolution at $1~{\rm GeV}/c$ is
$0.5\%$, and the $dE/dx$ resolution is $6\%$ for electrons
from Bhabha scattering. The EMC measures photon energies with a
resolution of $2.5\%$ ($5\%$) at $1$~GeV in the barrel (end cap)
region. The time resolution in the TOF barrel region is 68~ps, while
that in the end cap region is 110~ps. The end cap TOF
system was upgraded in 2015 using multi-gap resistive plate chamber
technology, providing a time resolution of
60~ps~\cite{etof,etof2,etof3}.  
About 87\% of the data used were collected after this upgrade.  

Simulated data samples produced with the {\sc
geant4}-based~\cite{geant4} MC package BOOST~\cite{bes:boost}, which
includes the geometric and material description of the BESIII detector~\cite{geo2,detvis} and the detector responses, are used to determine detection efficiencies
and to estimate backgrounds. The simulation models the beam
energy spread and initial state radiation (ISR) in the $e^+e^-$
annihilations with the generator {\sc
kkmc}~\cite{ref:kkmc,ref:kkmc2}. 
The inclusive MC sample includes the production of open charm
processes, the ISR production of vector charmonium(-like) states, and the continuum processes.
All particle decays are generated with {\sc
evtgen}~\cite{ref:evtgen} using BFs either taken from the Particle Data Group~(PDG)~\cite{pdg:2020}, when available, or otherwise estimated with {\sc lundcharm}~\cite{ref:lundcharm, ref:lundcharm2}. Final state radiation~(FSR)
from charged final state particles is incorporated using the {\sc
photos} package~\cite{photos}.
For the MC production of the $e^+e^-\rightarrow \Lambda_c^+\Lambda_c^-$ events, the cross section line-shape from BESIII measurements is taken into account.  The signal decay processes include an incoherent sum of intermediate state resonances, while for the reference mode $\Lambda_{c}^{+}\rightarrow \Sigma^+ \pi^+ \pi^-$, a partial-wave analysis is performed to obtain the amplitude model.

\section{EVENT SELECTION AND DATA ANALYSIS}
\label{sec:analysis}
\hspace{1.5em} 

In this work, both the signal and reference $\Lambda_c^+$ decays are fully reconstructed.  
Charged tracks detected in the MDC are required to be within a polar angle ($\theta$) range of $|\rm{cos\theta}|<0.93$, where $\theta$ is defined with respect to the $z$ axis, which is the symmetry axis of the MDC. The distance of closest approach to the interaction point (IP) must be less than 10\,cm along the $z$ axis and less than 1\,cm in the transverse plane.  

Particle identification~(PID) for charged tracks combines measurements of the energy deposited in the MDC~(d$E$/d$x$) and the flight time in the TOF to form likelihoods $\mathcal{L}(h)~(h=p,K,\pi)$ for each hadron $h$ hypothesis.
Tracks are identified as protons when the proton hypothesis has the greatest likelihood ($\mathcal{L}(p)>\mathcal{L}(K)$ and $\mathcal{L}(p)>\mathcal{L}(\pi)$), while charged kaons and pions are identified by comparing the likelihoods for the kaon and pion hypotheses, $\mathcal{L}(K)>\mathcal{L}(\pi)$ and $\mathcal{L}(\pi)>\mathcal{L}(K)$, respectively.

   Photon candidates are reconstructed using showers in the EMC.  The deposited energy of each shower must be more than 25~MeV in the barrel region ($|\cos \theta|< 0.80$) and more than 50~MeV in the end cap region ($0.86 <|\cos \theta|< 0.92$). To exclude showers that originate from charged tracks, the angle subtended by the EMC shower and the position of the closest charged track at the EMC must be greater than 10 degrees as measured from the IP. To suppress electronic noise and showers unrelated to the event, the difference between the EMC time and the event start time is required to be within [0, 700]\,ns.

All $\gamma\gamma$ combinations are considered as $\pi^{0}$ candidates,
 and the reconstructed mass $M(\gamma\gamma)$ is required to fall in the range of  
 $0.115$~GeV/$c^{2}<M(\gamma\gamma)<0.150$~GeV/$c^{2}$. A kinematic fit is performed to constrain the $\gamma\gamma$ invariant mass to the known $\pi^0$ mass~\cite{pdg:2020}, 
and candidates with the fit quality $\chi^2<200$ are retained.

The $\Sigma^+$ candidates are reconstructed from the combinations of $p\pi^0$ with an invariant mass in the range of $1.174$~GeV/$c^2< M(p\pi^0)< 1.200$~GeV/$c^2$. This requirement corresponds to approximately $\pm3\sigma$ (standard deviations) around the known $\Sigma^+$ mass~\cite{pdg:2020}. To reject the possible backgrounds for $\Lambda^+_c\to  \Sigma^+ K^+\pi^-$ and $\Sigma^+ \pi^+ \pi^-$ modes including $\Lambda \rightarrow p\pi^-$ in the final states, $M(p\pi^-)$ is required to be outside the range (1.11, 1.12) GeV/$c^2$. In addition, to remove the $K^{0}_{S}$ decays in the mode $\Lambda_{c}^{+}\to\Sigma^+ \pi^+ \pi^-$, events with $M(\pi^+\pi^-)$ in the range (0.48, 0.52)~GeV/$c^2$ are vetoed.  

\begin{figure}[!hbt]
\centering
\subfigure[]
{
	\includegraphics[width=0.3\paperwidth]{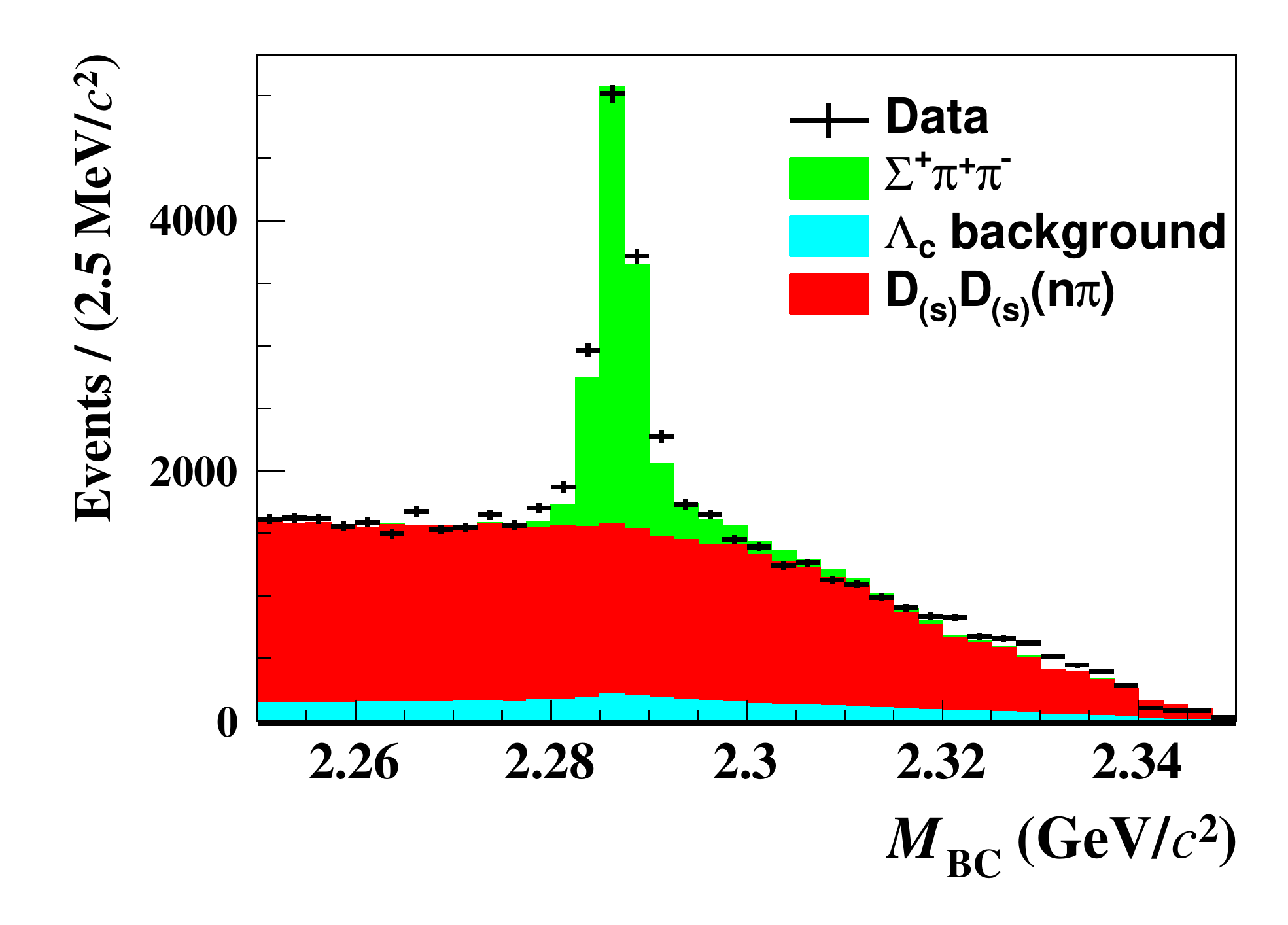}
}
\hspace{1pt}
\subfigure[]
{
	\includegraphics[width=0.3\paperwidth]{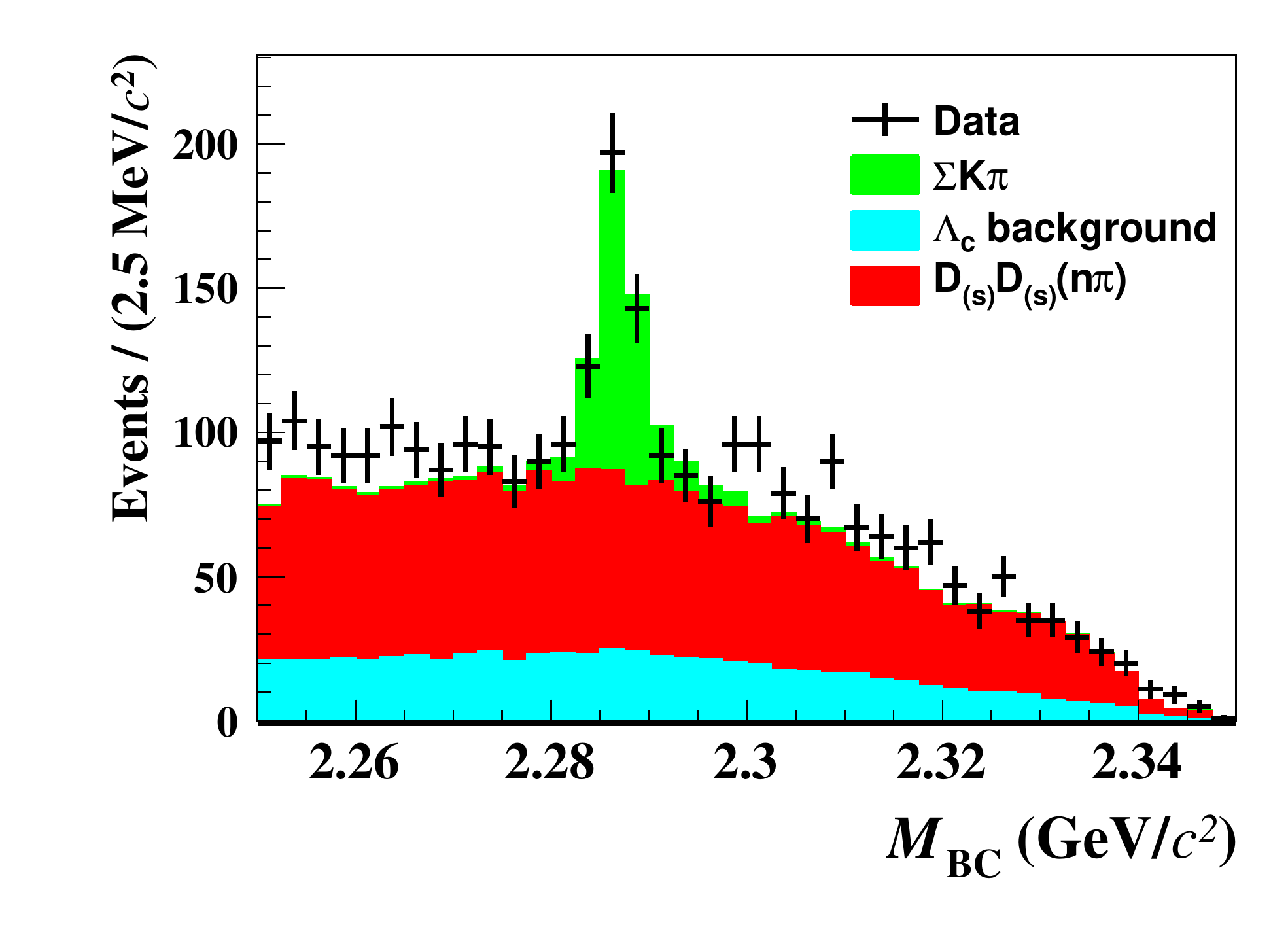}
}
\hspace{1pt}
\subfigure[]
{
	\includegraphics[width=0.3\paperwidth]{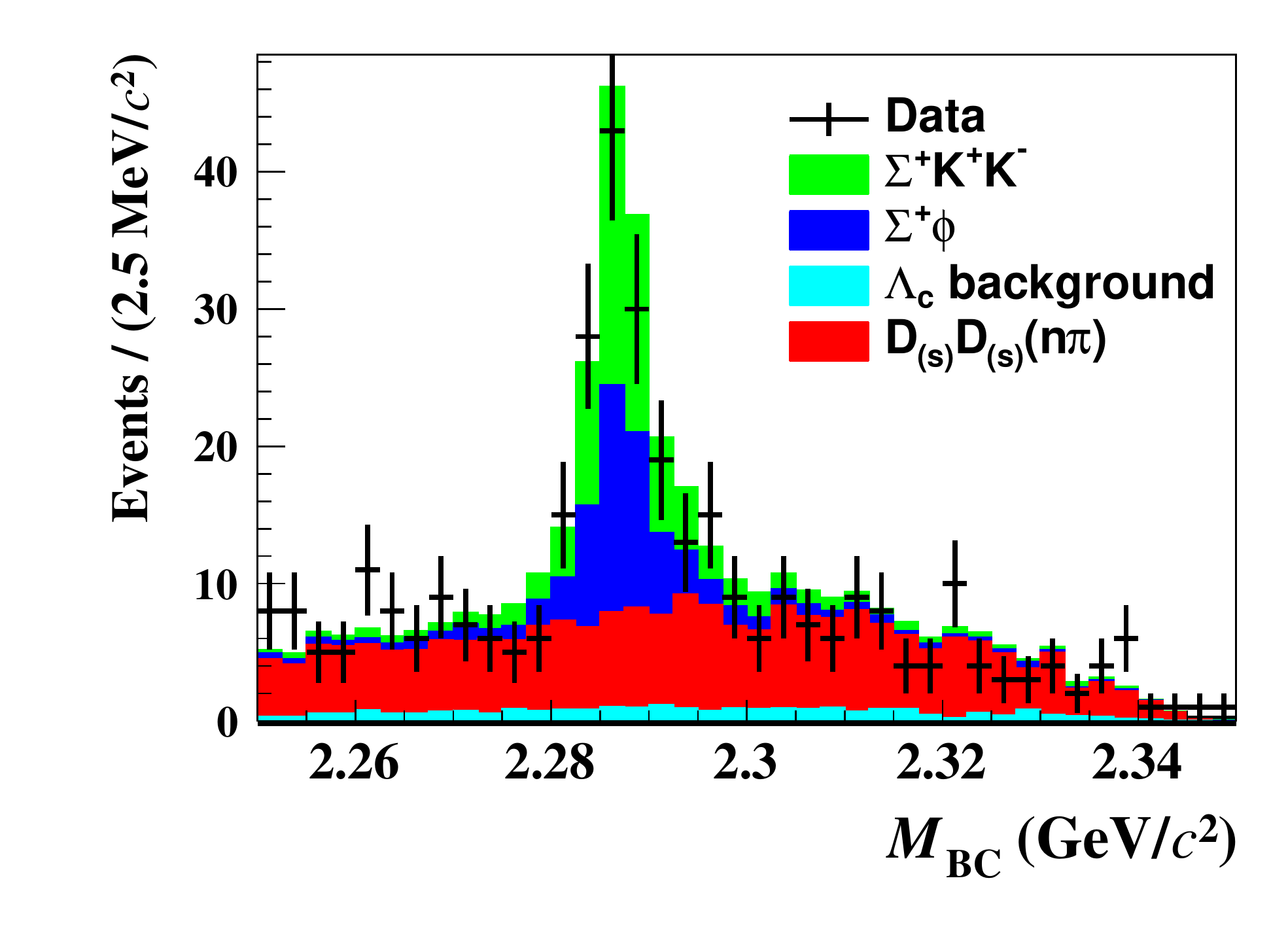}
}
\hspace{1pt}
\subfigure[]
{
	\includegraphics[width=0.3\paperwidth]{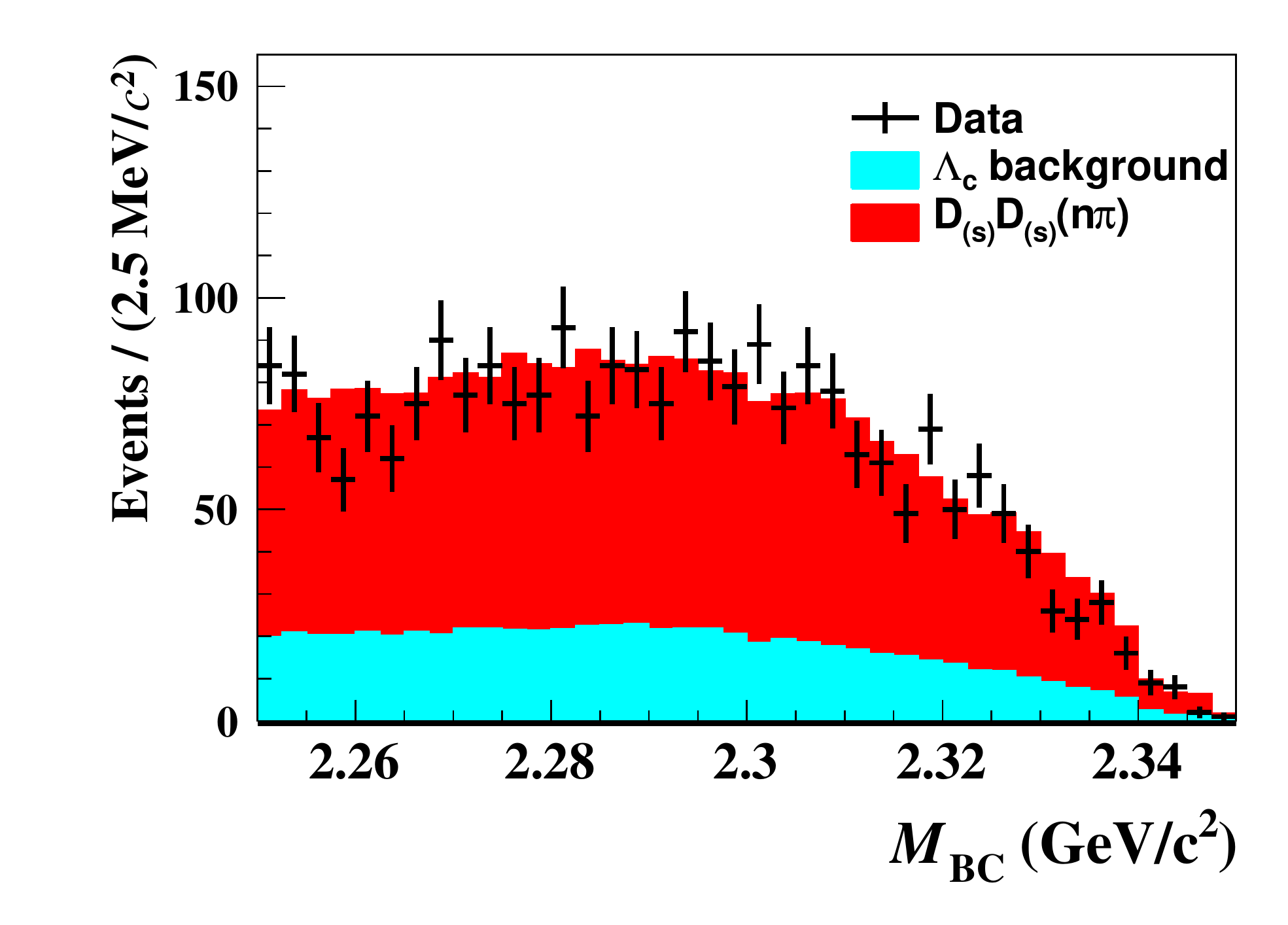}
}
\hspace{1pt}
\caption{The $M_{\rm BC}$ distributions of candidates for (a) $\Lambda_{c}^{+}\rightarrow\Sigma^{+} \pi^{+} \pi^{-}$, (b)$\Lambda_{c}^{+}\rightarrow\Sigma^{+} K^{+} \pi^{-}$,  (c)$\Lambda_{c}^{+}\rightarrow\Sigma^{+} K^{+} K^{-}$ and $\Lambda_{c}^{+}\rightarrow\Sigma^{+}\phi$, and (d)$\Lambda_{c}^{+}\rightarrow\Sigma^{+} K^{+} \pi^{-}\pi^{0}$  in data and inclusive MC samples. The black points with error bars are from the combined data at all seven energy points, the green and blue histograms are signal processes. The cyan and red histograms are the $\Lambda_c^+$ and open charm backgrounds, respectively.}
\label{fig:M_BC_bkg}
\end{figure}

To improve the signal purity, the energy difference $\Delta E = E_{\rm cand} - E_{\rm beam}$ for $\Lambda_c^+$ candidates are required to satisfy a mode-dependent $\Delta E$ requirement shown in Table~\ref{tab:DeltaE}.  These ranges are obtained by optimizing signal yield significance with the inclusive MC sample. Here $E_{\rm cand}$ is the total reconstructed energy of the $\Lambda_c^+$ candidate and $E_{\rm beam}$ is the beam energy. Only one candidate with the minimal $|\Delta E|$ is accepted. The $\Lambda^+_c$ signal is identified using the beam constrained mass $M_{\rm BC} = \sqrt{E_{\rm beam}^2/c^4 - p^2/c^2}$, where $p$ is the measured $\Lambda_c^+$ momentum in the center-of-mass system of the $e^+e^-$ collision.  After the event selection, there is no obvious peaking background in the $M_{\rm BC}$ distribution for each tag mode, as shown in Figure~\ref{fig:M_BC_bkg}.

\begin{table}[!htbp]
\footnotesize
   \caption{Requirements on $\Delta E$ for different $\Lambda_{c}^+$ decay modes.  }
   \begin{center}
   \begin{tabular}{c|c}
	\hline
	Decay mode & Requirement~(GeV) \\
	\hline
	\hline

	$\Sigma^{+} K^{+} K^{-}$ & $-0.017<\Delta E <0.008$    \\
	$\Sigma^{+} K^{+} \pi^{-}$ & $-0.014<\Delta E <0.008$\\
	$\Sigma^{+} K^{+} \pi^{-} \pi^{0}$ & $-0.028<\Delta E <0.012$  \\
	$\Sigma^{+} \pi^{+} \pi^{-}$ & $-0.040<\Delta E <0.032$\\
	\hline
	\hline
   \end{tabular}
    \end{center}
	 \label{tab:DeltaE}
\end{table}

\section{DETERMINATION OF THE BRANCHING FRACTIONS}
\label{sec:bf}
\hspace{1.5em}

 The RBF between the signal and reference modes is calculated with  

  \begin{equation}\label{eq1}
{RBF}_{ij} = \frac{\mathcal{B}_{i}}{\mathcal{B}_{j}} = \frac{N_{i}\cdot\varepsilon_{j}\cdot\mathcal{B}_{{\rm inter}}^{j}}{N_{j}\cdot\varepsilon_{i}\cdot\mathcal{B}_{{ \rm inter}}^{i}}.  
  \end{equation}
where $\varepsilon_{i}$ and $\varepsilon_{j}$ are the detection efficiencies, $N_{i}$ and $N_{j}$ are the signal yields of the signal mode $i$ and the reference mode $j$, respectively. The $\mathcal{B}_{\rm{inter}}$ is the product of the BFs of the intermediate states ($\Sigma^+\rightarrow p\pi^0$, $\pi^0 \rightarrow \gamma \gamma$, and $\phi\to K^{+} K^{-}$ for $\Lambda_{c}^{+}\rightarrow\Sigma^{+}\phi$), and the charge-conjugate channel is included in the simulation. The observables in Equation~\eqref{eq1} are determined as follows. 

\begin{itemize}
        \item \emph{Reference mode $\Lambda_{c}^{+}\rightarrow\Sigma^{+} \pi^{+} \pi^{-}$}

To obtain the yield of the reference mode, an unbinned maximum likelihood (UML) fit is performed on the $M_{\rm BC}$ distribution separately at each energy point for data. In each fit, the signal shape is described by the MC simulated shape convolved with a Gaussian function with floating mean and width, in order to take into account the resolution difference between data and MC simulation. The distribution of the combinatorial backgrounds is modeled with an ARGUS function~\cite{ARGUS:1990hfq}
\begin{equation}
	f^\mathrm{ARGUS} \propto M_{\mathrm{BC}} \sqrt{1-\left(\frac{M_{\mathrm{BC}}\cdot c^2}{E_\mathrm{beam}}\right)^{2}} \,\, e^{a\left(1-\frac{M_{\mathrm{BC}}\cdot c^2}{E_\mathrm{beam}}\right)^{2}},
	\label{eq:argus}
\end{equation}
where the parameter $a$ is free in the fit. 
The fitted yields of the reference mode at different energy points are given in Table~\ref{tab:yields}. In addition, the detection efficiencies are estimated according to MC simulations, as listed in Table~\ref{tab:efficiency}.
The total fit to data summing over all the energy points is shown in Figure~\ref{fig:pipi}.

\begin{table}[!tp]
   \caption{Signal yields in data for various decay modes, where the uncertainties are statistical only. $\Lambda^+_c\to\Sigma^{+} K^{+} \pi^{-} \pi^{0}$ has no significant signal observed. }
     \scriptsize
\begin{center}
   \begin{tabular}{cccccccc}

	\hline
	Decay mode &4.600~GeV &4.612~GeV &4.628~GeV &4.641~GeV &4.661~GeV &4.682~GeV &4.699~GeV\\
	\hline
	\hline
      $\Sigma^{+} \pi^{+} \pi^{-}$ &1123$\pm$47&200$\pm$21&1003$\pm$46&1026$\pm$48&1025$\pm$48&3132$\pm$85&942$\pm$47\\
      
	$\Sigma^{+} K^{+} K^{-}$ &\ \ 14.1$\pm$1.5&\ \  2.3$\pm$0.2 &\ \ 11.7$\pm$1.3& \ \ 13.0$\pm$1.4 &\ \ 12.1$\pm$1.4&\ \  43.0$\pm$4.5&\   14.1$\pm$1.4\\
	
	$\Sigma^{+}\phi$ &\ \ 14.3$\pm$2.8&\ \  2.4$\pm$0.5& \ \ 12.7$\pm$2.5&\ \  13.2$\pm$2.7&\ \  14.2$\pm$2.7&\ \  47.2$\pm$8.9&\ 14.9$\pm$2.9 \\
	
	$\Sigma^{+} K^{+} K^{-}$(non-$\phi$) & \ \  \  \ 9.2$\pm$1.7&\ \ 1.6$\pm$0.3& \ \ \ \ 8.1$\pm$1.5&\ \  \ \ 8.6$\pm$1.6&\ \  \ \  9.0$\pm$1.6&\ \  29.0$\pm$4.0& \ 9.1$\pm$1.6\\
	
	$\Sigma^{+} K^{+} \pi^{-}$ &\ \ 30.5$\pm$3.6&\ \ 5.8$\pm$0.7& \ \ 29.3$\pm$3.5&\ \  29.6$\pm$3.5&\ \  26.1$\pm$3.4&\ \  80.3$\pm$9.5&  \ 22.3$\pm$2.6 \\
	\hline
	\hline
   \end{tabular}
    \end{center}
	 \label{tab:yields}
\end{table}

\begin{table}[!tp]
   \caption{Detection efficiencies (in unit of \%) for various decay modes, where the uncertainties are statistical only. The efficiencies do not include the BF of the sequential decay of $\pi^{0}$, $\phi$ or $\Sigma^{+}$. }
  \scriptsize
\begin{center}
   \begin{tabular}{cccccccc}

	\hline
	Decay mode &4.600~GeV &4.612~GeV &4.628~GeV &4.641~GeV &4.661~GeV &4.682~GeV &4.699~GeV \\
	\hline
	\hline
       $\Sigma^+ \pi^+ \pi^-$&26.12$\pm$0.05 &25.21$\pm$0.05&24.90$\pm$0.05&24.96$\pm$0.05&24.78$\pm$0.05&24.44$\pm$0.05&24.23$\pm$0.05\\
	$\Sigma^{+} K^{+} K^{-}$ & \ \ 4.22$\pm$0.02& \ \  3.92$\pm$0.02& \ \ 3.98$\pm$0.02& \ \ 4.11$\pm$0.02& \ \  4.30$\pm$0.02& \  \ 4.53$\pm$0.02& \  \ 5.19$\pm$0.02\\
	$\Sigma^{+}\phi$ & \  \ 3.66$\pm$0.02& \ \ 3.38$\pm$0.02& \ \  3.46$\pm$0.02& \ \  3.55$\pm$0.02& \ \  3.75$\pm$0.02& \ \  4.02$\pm$0.02& \  \ 4.19$\pm$0.02 \\
	$\Sigma^{+} K^{+} K^{-}$(non-$\phi$) &\  \ 4.81$\pm$0.02&\  \ 4.50$\pm$0.02&\  \ 4.53$\pm$0.02&\  \ 4.70$\pm$0.02&\  \ 4.88$\pm$0.02&\  \ 5.07$\pm$0.02&\  \ 6.24$\pm$0.02\\
	$\Sigma^{+} K^{+} \pi^{-}$ &16.30$\pm$0.04&15.78$\pm$0.04&15.20$\pm$0.04&15.06$\pm$0.04&14.56$\pm$0.04&14.75$\pm$0.04&14.20$\pm$0.04 \\
	$\Sigma^{+} K^{+} \pi^{-} \pi^{0}$ & \  \ 5.16$\pm$0.02& \ \  4.72$\pm$0.02& \  \ 4.55$\pm$0.02& \ \  4.59$\pm$0.02& \ \ 4.73$\pm$0.02& \ \ 4.72$\pm$0.02& \ \ 4.74$\pm$0.02 \\
       
	\hline
	\hline
   \end{tabular}
    \end{center}
	 \label{tab:efficiency}
\end{table}

\begin{figure}[!hbt]
\centering
\subfigure[]
{
	\includegraphics[width=0.3\paperwidth]{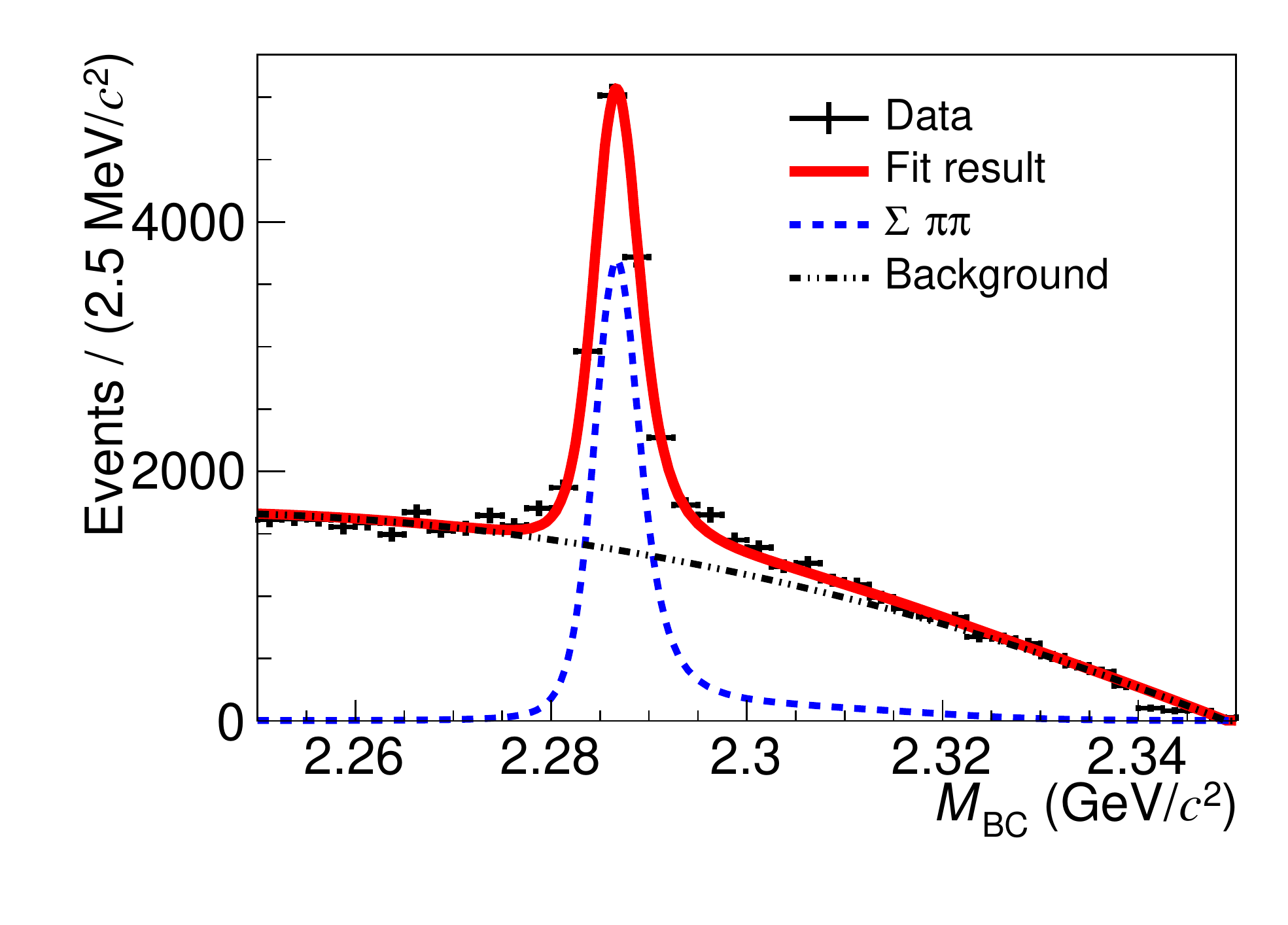}
	\label{fig:pipi}
}
\hspace{1pt}
\subfigure[]
{
	\includegraphics[width=0.3\paperwidth]{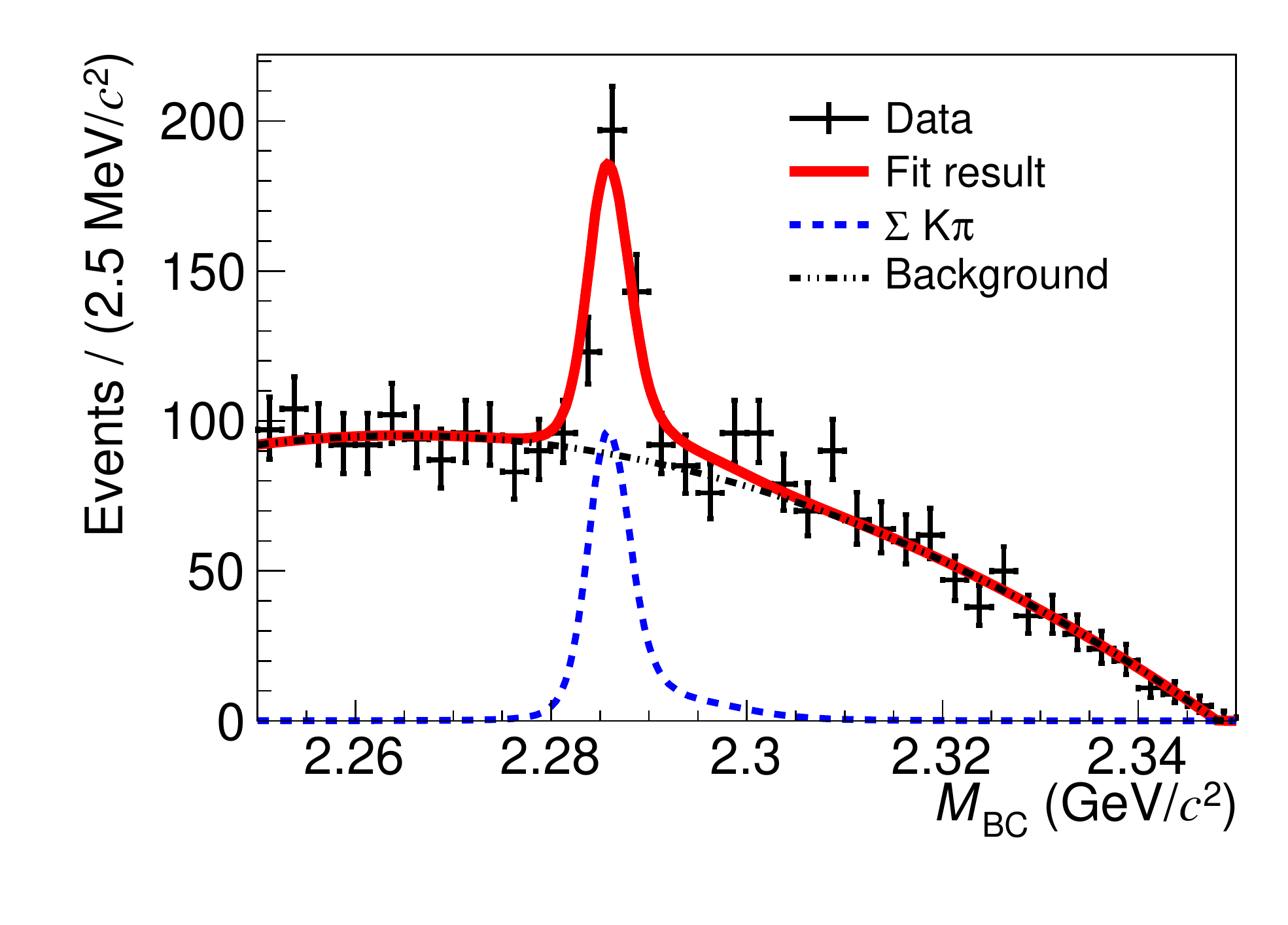}
	\label{fig:kpi}}
\hspace{1pt}
\subfigure[]
{
	\includegraphics[width=0.3\paperwidth]{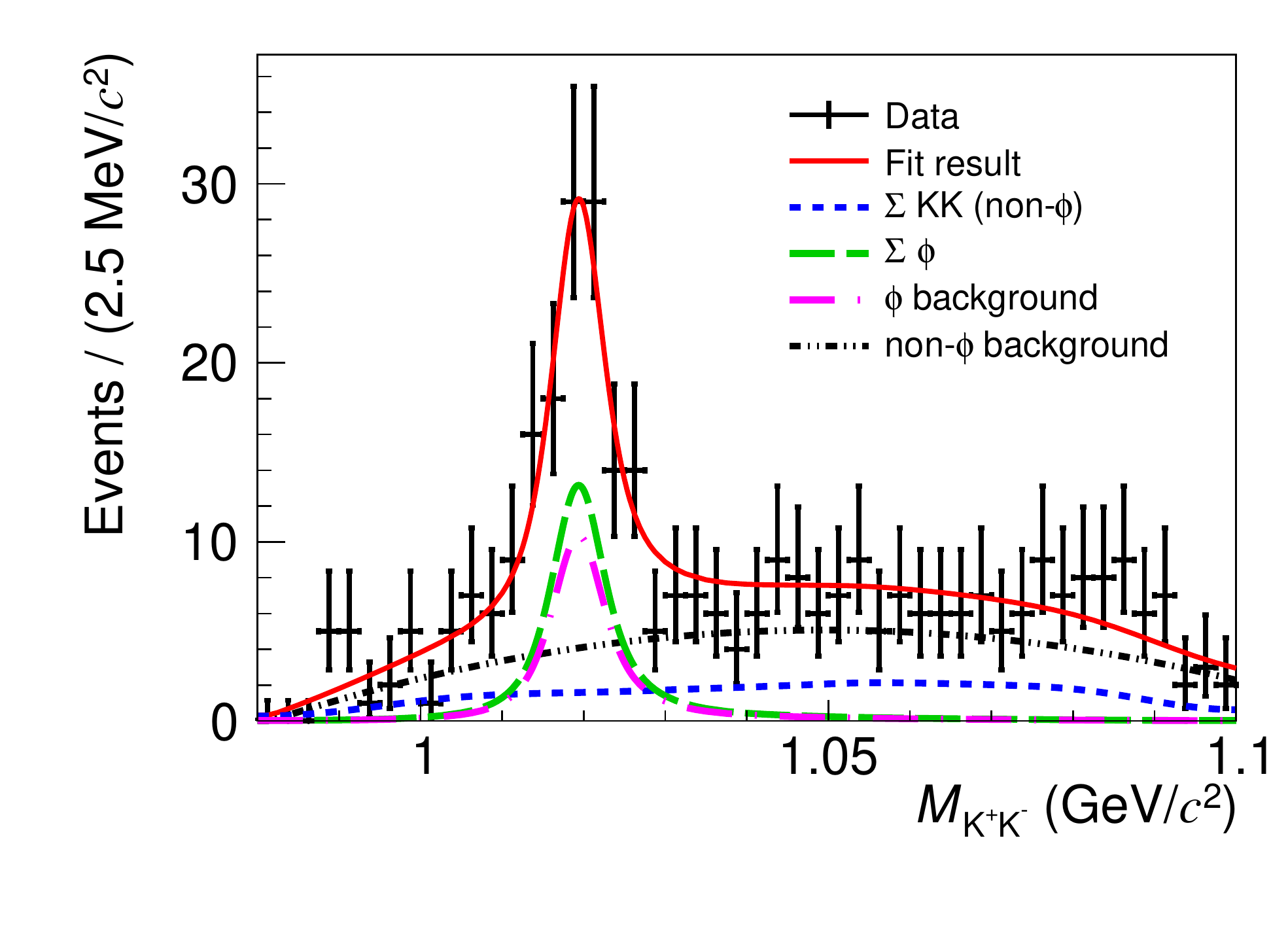}
	\label{fig:kk1}
}
\hspace{1pt}
\subfigure[]
{
	\includegraphics[width=0.3\paperwidth]{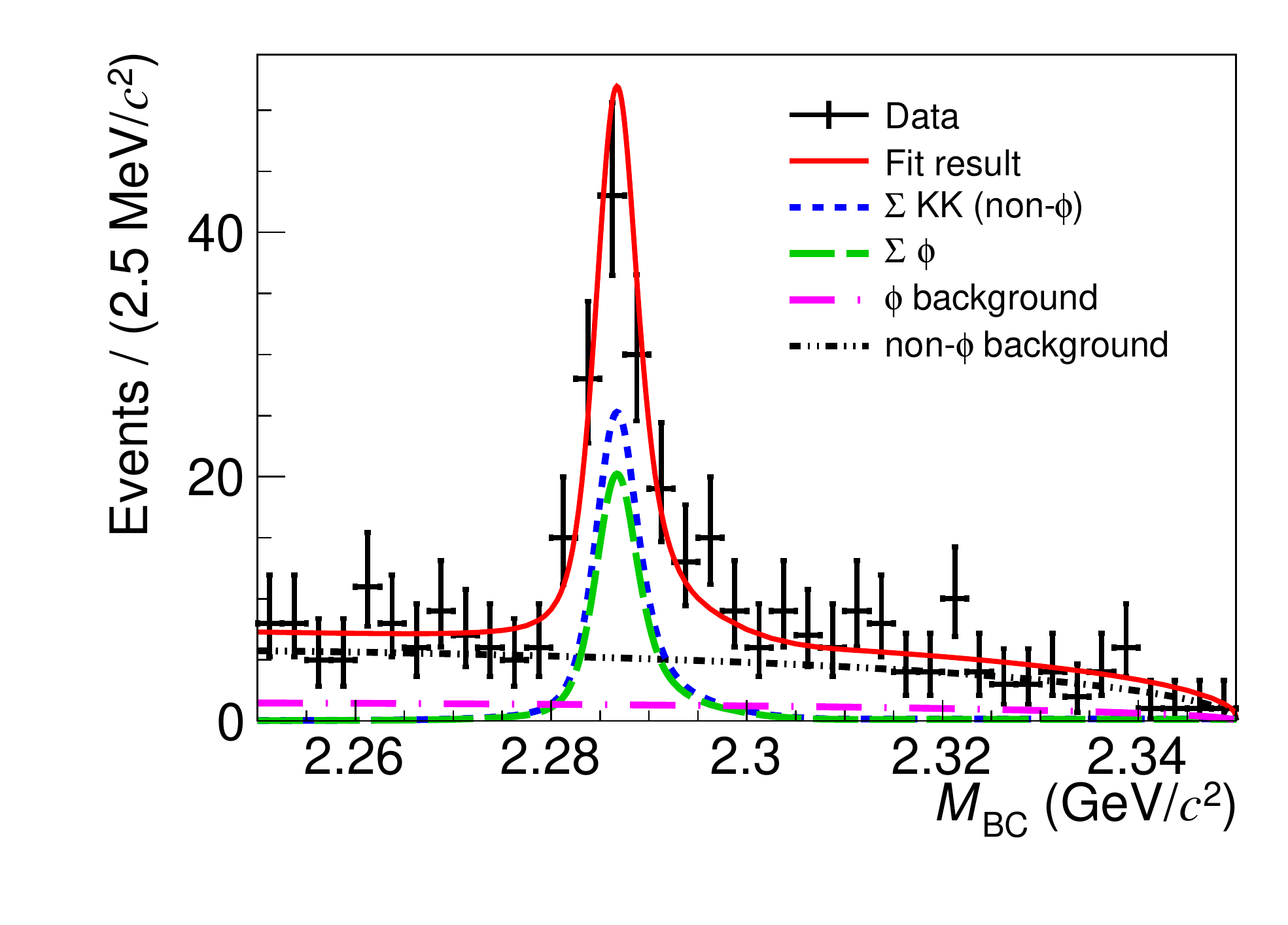}
	\label{fig:kk2}
}
\label{fig:up1}
\hspace{1pt}
\subfigure[]
{
	\includegraphics[width=0.3\paperwidth]{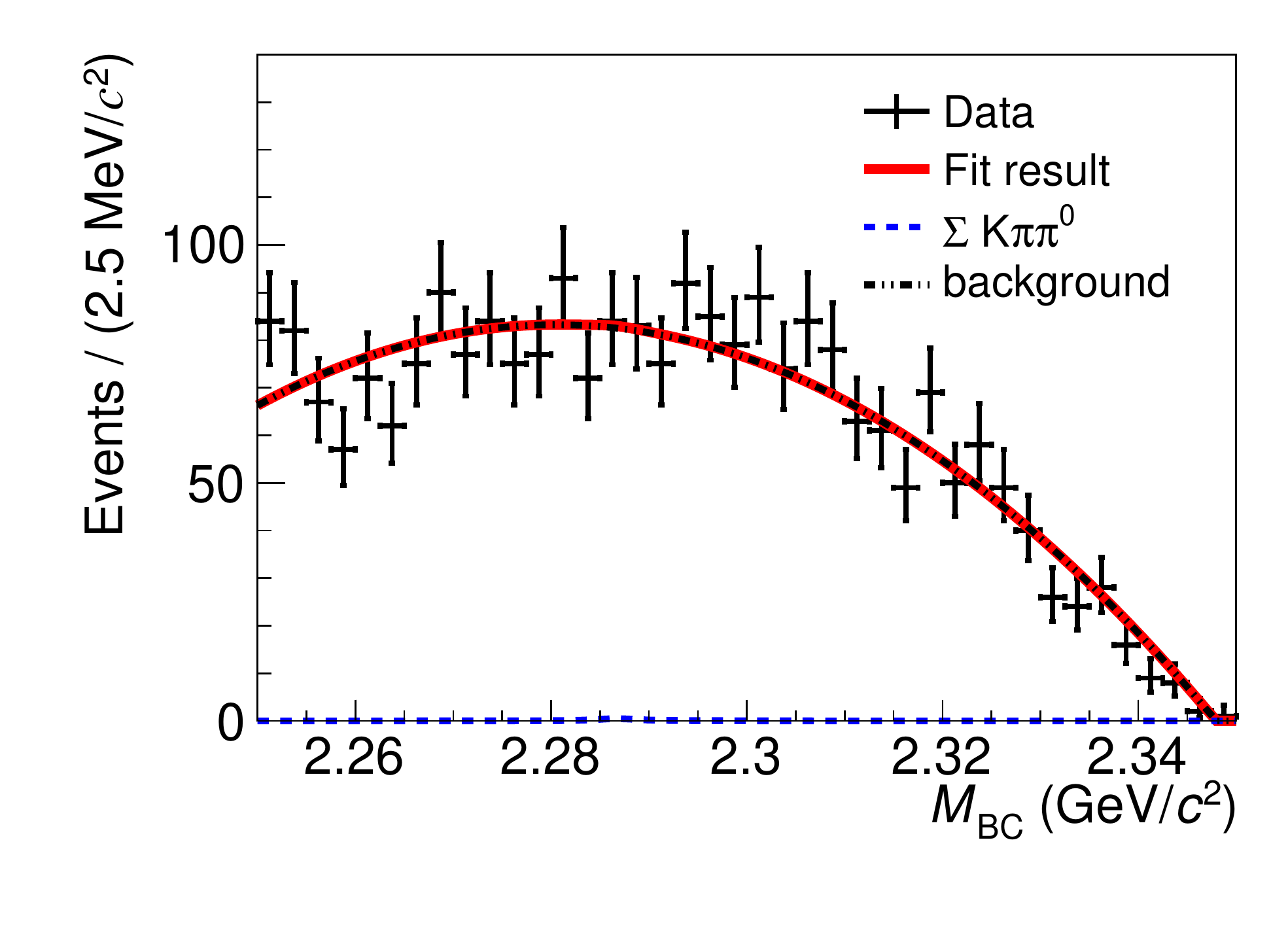}
	\label{fig:kpipi0}
}

\caption{Fit results for different decay modes. The plots show the $M_{\rm BC}$ distributions of $\Lambda_{c}^{+}\rightarrow\Sigma^{+} \pi^{+} \pi^{-}$ (a), $\Lambda_{c}^{+}\rightarrow\Sigma^{+} K^{+} \pi^{-} $ (b) and $\Lambda_{c}^{+}\rightarrow\Sigma^{+} K^{+} \pi^{-} \pi^{0}$ (e) processes. Other plots (c) and (d) show the distributions of $M_{\rm BC}$ and $M_{K^+K^-}$, respectively,  for the projections of the 2-D fit used to separate the $\Lambda_{c}^{+}\rightarrow\Sigma^{+}\phi$ and $\Lambda_{c}^{+}\rightarrow\Sigma^{+} K^{+} K^{-}$(non-$\phi$)  processes.  The points with error bars are combined from data at all energy points, the red curves are the overall fit result, the blue and green dashed curves are the signal shapes, the black and pink dashed curves are the background shapes. }
\label{fig:M_BC_fit}
\end{figure}

\item \emph{Signal mode $\Lambda_{c}^{+}\rightarrow\Sigma^{+} K^{+} \pi^{-}$}

The simultaneous UML fit is performed using seven energy point data samples for signal mode, to obtain a more precise result.  
For this purpose, a common RBF is fitted at various energy points, where the signal yields can be derived from Equation~\ref{eq1} with the input yields of the reference mode in Table~\ref{tab:yields} and the detection efficiencies in Table~\ref{tab:efficiency}. The uncertainties of the input values are taken into account in systematic uncertainties, as listed in Table~\ref{tab:sys}. 
In the fit, the signal shape is extracted from the corresponding signal MC sample and convolved with a Gaussian function with floating mean and width. The background shape is described by an ARGUS function in Equation~\ref{eq:argus}. 
The summed fit to data from all of the energy points is shown in Figure~\ref{fig:kpi}.
The RBF results obtained from the fit are given in Table~\ref{tab:sum}, and the signal yields at different energy points are calculated, as listed in Table~\ref{tab:yields}.


\item \emph{Signal mode $\Lambda_{c}^{+}\rightarrow\Sigma^{+} K^{+} K^{-}$}

To separate the $\phi$ contribution for $\Sigma^{+} K^{+} K^{-}$ mode, a two-dimensional simultaneous UML fit is performed on the $M_{\rm BC}$ vs the $M_{K^+K^-}$ distributions, in which common RBF values are estimated. Four components ($\Sigma^+ \phi$, $\Sigma^+ K^+ K^-$(non-$\phi$), $\phi$ background and non-$\phi$ background) are considered in this fit, as shown in Figures~\ref{fig:kk1} and \ref{fig:kk2}.  All the signal shapes are extracted from MC samples and convolved with a Gaussian resolution function with floating mean and width. In the $M_{\rm BC}$ distribution, the combinatorial background is  described using the ARGUS function. In the $M_{K^+K^-}$ distribution, we use a second-order Chebyshev function to describe the background of the non-$\phi$ process and use the $\phi$ line-shape distribution extracted from the $\Lambda^{+}_c \to \Sigma^+ \phi$ signal MC sample to describe the $\phi$ background.

\item \emph{Signal mode $\Lambda_{c}^{+}\rightarrow\Sigma^{+} K^{+} \pi^{-}\pi^0$}

The simultaneous UML fit is also performed for $\Lambda_{c}^{+}\rightarrow\Sigma^{+} K^{+} \pi^{-}\pi^0$, as shown in Figure~\ref{fig:kpipi0}. Since there is no significant signal observed, the upper limit on the BF of this decay is estimated with a likelihood scan method which takes into account the systematic uncertainties as follows
 \begin{equation}\label{eq:upper}
L(\rm{BF}) = \int_{-1}^{1} \emph{L}^{\rm{stat}} \, [(1+\Delta) \, \rm{BF}] \, \rm exp(-\frac{\Delta^2}{2\sigma^2_{syst}}) \, d\Delta.
 \end{equation}
Here, $L(\rm{BF})$ is the likelihood expression of BF,  $L^{\rm{stat}}$ is the statistical likelihood given by the data without considering the systematic uncertainties when taking the nominal BF obtained, $\Delta$ is the relative deviation of the estimated BF from the nominal value and $\sigma_{\rm{syst}}$ is the total systematic uncertainty given in Table~\ref{tab:sys}.
The likelihood curve calculated according to Equation~\ref{eq:upper} is shown in Figure~\ref{fig:upper}. The upper limit on the BF of $\Lambda^+_c\to\Sigma^+ \pi^- \pi^+ \pi^0$ mode at the 90\% confidence level (CL) is estimated to be 0.11\% by integrating the likelihood curve in the physical region and considering the BF of the reference mode and systematic uncertainties.

\vspace{-0.0cm}
\begin{figure*}[!htbp]
\centering
\includegraphics[width=0.5\textwidth]{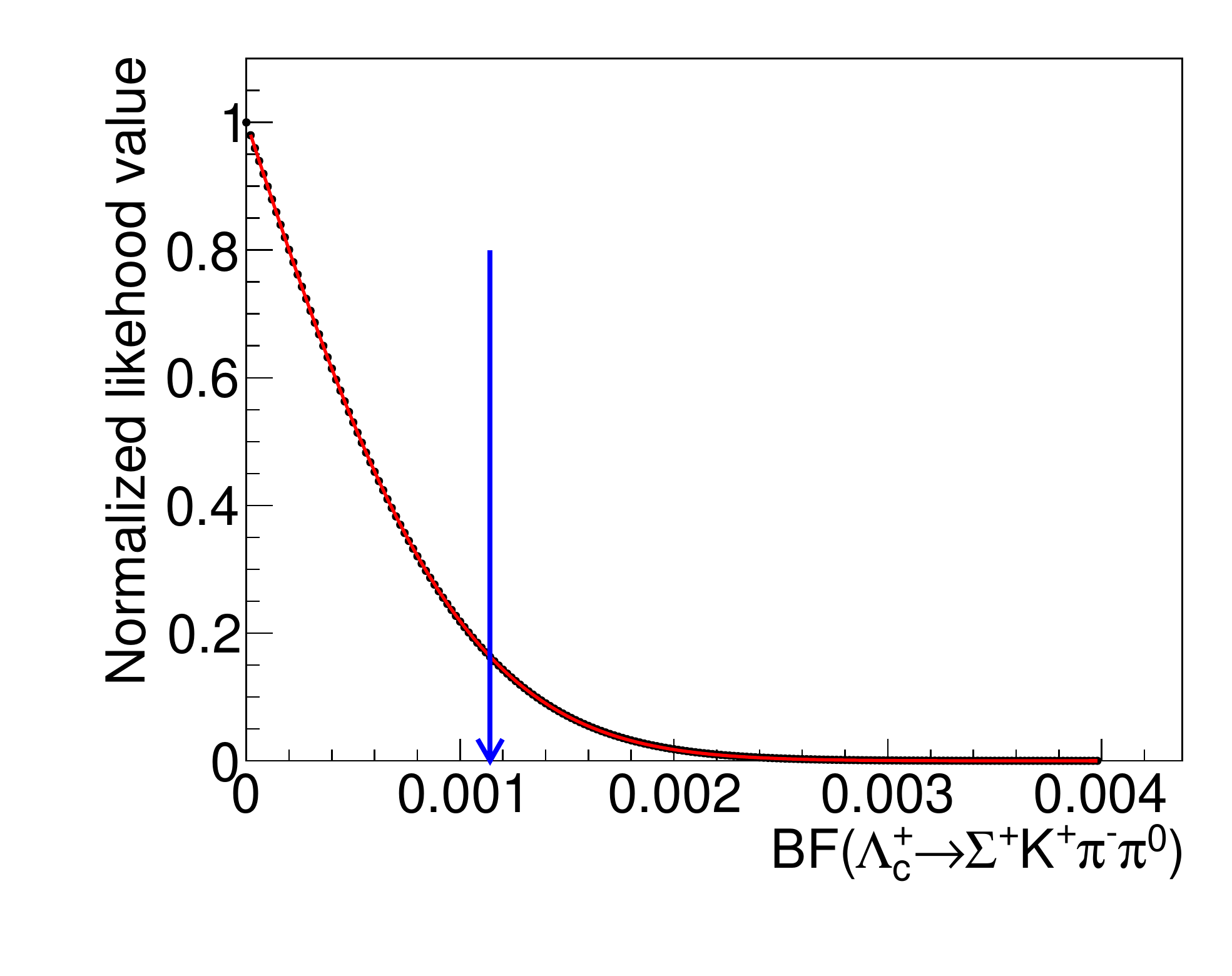}
\caption{The normalized likelihood value versus the BF of $\Lambda^+_c\to \Sigma^+K^+\pi^-\pi^0$ after incorporating the systematic uncertainty.  The black points are the initial curve, the red curve is the result with the systematic uncertainty, and the blue arrow points to the upper limit on the BF at the 90$\%$ CL.}
\label{fig:upper}
\end{figure*}
\vspace{-0.0cm}


\end{itemize}

\section{SYSTEMATIC UNCERTAINTY}
\label{sec:systematic}
\hspace{1.5em}

The sources of systematic uncertainties in the RBF measurements are summarized in Table~\ref{tab:sys}. The systematic uncertainties of MC statistics is $0.1\%$ for each channel and can be neglected.
The systematic uncertainties in $\Sigma^+$ reconstruction in both signal and reference modes are canceled in the RBF calculation. The systematic uncertainties of tracking efficiency for charged tracks are also canceled. For each signal decay, the square root of the quadratic sum of all the uncertainties is taken as the total systematic uncertainty.

\begin{table}[!hbtp] 
	 \caption{Relative systematic uncertainties in the RBF measurements, in \%. }
\begin{center}
\scriptsize
   \begin{tabular}{cccccc}
	 \hline \hline
	 
   Source        &  $\Sigma^{+}  K^{+} K^{-}$ & $\Sigma^{+} K^{+} \pi^{-}$& $\Sigma^{+} K^{+} \pi^{-}\pi^{0}$& $\Sigma^{+}\phi$ & $\Sigma^{+}  K^{+} K^{-}_{{\rm non-}\phi}$
  
	 \\

		    \hline
      $K^{\pm}$  PID         &  2.0 & 1.0 & 1.0& 2.0 & 2.0\\
       $\pi^{\pm}$   PID   &  2.0 & 1.0 & 1.0& 2.0 & 2.0\\
       $\pi^{0}$ reconstruction &          -& -&1.0& -&- \\
	 $\Delta E$ requirement &  1.4   & 3.0  &  3.5 & 1.4& 1.4 \\
	 $M_{\rm{BC}}$ fit&                     3.3   & 3.0&  5.9 & 6.1& 3.0 \\

	 Signal model                       & 1.8 & 2.9 & - & 1.0&0.7\\ Reference mode &1.9&1.9&1.9&1.9&1.9\\
	 ${\mathcal B}_{\rm{inter}}$ &                       0.6  & 0.6   & 0.6 &1.2 &0.6\\

	  \hline
	 Total          &                          5.3  &    5.7    &  7.4 &7.3 & 4.8\\
	 \hline\hline
	 \end{tabular}

	 \label{tab:sys}
	 \end{center}
	  \end{table}

\begin{itemize}
    \item \emph{PID and $\pi^0$ reconstruction efficiencies}.
     The uncertainties associated with PID efficiencies are estimated to be 1.0\% for each charged track by studying the control samples of $e^+e^-\rightarrow \pi^+\pi^-\pi^+\pi^-$, $K^+K^-\pi^+\pi^-$ and $p\bar{p}\pi^+\pi^-$ based on data taken at energies above $\sqrt{s} = 4.0$~GeV.  However, the PID uncertainty for the $p(\bar{p})$ common to signal and reference modes cancels, as does that for any common $\pi^{\pm}$. The uncertainty of $\pi^0$ reconstruction efficiency is assigned to be 1.0\% per $\pi^0$ by studying the control sample of $e^+e^-\rightarrow \omega \pi^0$ process.

    \item \emph{$\Delta E$ requirement}.
    The resolution difference between data and MC simulation need to be considered. Thus a Gaussian function is used to smear MC sample and obtain a new efficiency. The changes in the efficiencies are assigned as the corresponding uncertainties of $\Delta E$. The resolution difference is extract by using signal MC shape smeared with the Gaussian function to fit the $\Delta E$ distribution in data.
    
    \item \emph{$M_{BC}$ fit}.
     The uncertainty in the $M_{\rm BC}$ fit is mainly due to free parameters of the Gaussian and ARGUS functions and the background description. The relevant systematic uncertainty is estimated with an alternative signal shape without smearing the Gaussian resolution function and an alternative background shape which varies the high-end cutoff of the ARGUS function by $\pm 0.005$ GeV/$c^2$. The largest result change is assigned as the systematic uncertainty. For $\Lambda_{c}^{+}\rightarrow\Sigma^{+}\phi$ process, due to that a two-dimensional maximum likelihood fit is performed on the $M_{\rm BC}$ and $M_{K^{+}K^{-}}$ distribution, the uncertainty of $M_{K^{+}K^{-}}$ fit is also considered by studying the difference of $\phi$ signal shape convolved with or without a Gaussian resolution function. In addition, the shape of the backgrounds from $\Lambda^+_{c}$ inclusive decays can not be different from the ARGUS function. To account for this discrepancy, an additional component of the $\Lambda^+_{c}$ decay backgrounds, which is extracted from the $\Lambda^+_{c}$ inclusive MC samples, is included in the $M_{\rm{BC}}$ fit. The resultant changes on the RBF are taken as the systematic uncertainties. For $\Lambda_{c}^{+}\rightarrow\Sigma^+K^+\pi^-\pi^{0}$ process, an alternative fit to data with $M_{\rm{BC}}$ larger than 2.27 GeV is implemented and the relative changes on the fitting results are taken into account as systematic uncertainties.  All the above items are summed over in quadrature.

    \item \emph{Signal model}.
     The influence of the assumed signal model on the BF measurement comes from the estimation of the signal efficiency.
Decay processes with intermediate resonances listed in the PDG \cite{pdg:2020} are included in the signal simulation.  Examples include $\Lambda_{c}^{+}\rightarrow\Sigma^{+}\phi$ and $\Lambda_{c}^{+}\rightarrow\Xi^{*0} K^{+}$ ($\Xi^{*0}\rightarrow \Sigma^{+} K^{-}$) processes for $\Lambda_{c}^{+}\rightarrow\Sigma^{+}  K^{+} K^{-}$ and $\Lambda_{c}^{+}\rightarrow\Sigma^{+} K^{*}(892)^0$ ($K^{*}(892)^0\rightarrow K^+ \pi^-$) for $\Lambda_{c}^{+}\rightarrow\Sigma^{+} K^{+} \pi^{-}$. In the nominal analysis, the BFs that are used in the generator are taken from the PDG \cite{pdg:2020}. Their effects on the new BF measurements are estimated by varying the input BF of the intermediate decay by $\pm 1\sigma$ in the generator. The relative efficiency deviation is taken as the uncertainty. 

\item \emph{Reference mode}.
For the reference mode $\Lambda_{c}^{+}\rightarrow\Sigma^{+}\pi^{+}\pi^{-}$, the TF-PWA~\cite{PWA} is used to perform the simple Partial Wave Analysis and consider the possible intermediate resonance states, In the estimation of systematic uncertainty, we remove the low significance state. The difference of efficiency (1.1\%) is taken as the systematic uncertainty. For the  statistical uncertainties of the fitted yields in the reference mode, a relative uncertainty (1.6\%) is transferred into the systematic uncertainty of the RBF. In total, a quadrature sum of the systematic uncertainty (1.9\%) is assigned.

\item \emph{BFs of the intermediate states}.
     The BFs of $\pi^{0}\rightarrow\gamma\gamma$, $\phi\rightarrow K^{+}K^{-}$ and $\Sigma^{+}\rightarrow p^{+}\pi^{0}$ are used as inputs in the baseline analysis, and their uncertainties~\cite{pdg:2020} are propagated as the systematic uncertainty.
     
        
\end{itemize}

\section{SUMMARY}
\label{sec:summary}
\hspace{1.5em}
Based on 4.5 fb$^{-1}$ data taken at $\sqrt{s}=4.600$ to 4.699 GeV with the BESIII detector at the BEPCII collider,  
the non-factorizable $W$-exchange-only processes, $\Lambda_{c}^{+}\rightarrow\Sigma^{+}\phi$, $\Lambda_{c}^{+}\rightarrow\Sigma^{+} K^{+} K^{-}$(non-$\phi$) and the non-factorizable $W$-emission processes $\Lambda_{c}^{+}\rightarrow\Sigma^{+} K^{+} \pi^{-}$ and $\Lambda_{c}^{+}\rightarrow\Sigma^{+} K^{+} \pi^{-}\pi^{0}$ have been studied. The BF of $\Lambda_{c}^{+}\rightarrow\Sigma^{+} K^{+} \pi^{-}\pi^{0}$ relative to $\Lambda_{c}^{+}\rightarrow\Sigma^{+} \pi^{+} \pi^{-}$ is measured for the first time. The precisions of BFs for other channels relative to $\Lambda_{c}^{+}\rightarrow\Sigma^{+} \pi^{+} \pi^{-}$ are improved. Combining with the world average $\mathcal{B}(\Lambda_{c}^{+}\rightarrow\Sigma^{+} \pi^{+} \pi^{-}) = (4.50\pm 0.25)\%$~\cite{pdg:2020}, the BFs of the aforementioned decays are obtained. Table~\ref{tab:vs} shows the comparison of our results with the PDG values~\cite{pdg:2020} and the Belle results~\cite{Belle:2001hyr}. The uncertainties of the BFs of  $\Lambda_{c}^{+}\rightarrow\Sigma^{+} K^{+} K^{-}$ and $\Lambda_{c}^{+}\rightarrow\Sigma^{+}\phi$ are comparable to those in references ~\cite{Belle:2001hyr,pdg:2020}.
 For $\Lambda_{c}^{+}\rightarrow\Sigma^{+} K^{+} \pi^{-}$, the precision of the BF is improved by a factor of two. The theoretical predictions of the BFs of $\Lambda_{c}^{+}\rightarrow\Sigma^{+} \phi$, $\Lambda_{c}^{+}\rightarrow\Sigma^{+} K^{+} K^{-}$ and $\Lambda_{c}^{+}\rightarrow\Sigma^{+} K^{+} \pi^{-}$ in reference~\cite{Cen:2019ims, Hsiao:2019yur} differ from our results by about $2\sigma$ and our results will be helpful to correct the theoretical model.
The combined results are essential to understand the non-factorizable $W$-exchange and $W$-emission contributions to the decays of $\Lambda_c^+$.
\begin{table}[!hbtp] 
\footnotesize
	 \caption{Comparison of our RBF and BF results with the Belle results~\cite{Belle:2001hyr} and the PDG values~\cite{pdg:2020} (in unit of \%). Except for the mode $\Sigma^{+} K^{+} \pi^{-}\pi^{0}$, in all our RBF results, the first uncertainties are statistical, and the second are systematic. The third uncertainties of our BF results are from external input of the branching fraction of $\Lambda_{c}^{+}\rightarrow \Sigma^+ \pi^+ \pi^-$.
     }
   \begin{center}
   \begin{tabular}{ccccc}
	 \hline \hline
	 Decay mode  &RBF~(This work) &RBF~(Belle) &BF~(This work) &BF~(PDG)  \\  
	 \hline
	 $\Sigma^{+} K^{+} K^{-}$       &  $8.38\pm0.93\pm0.44$    &$7.6\pm0.7\pm0.9$& $0.377\pm0.042\pm0.020\pm0.021$  & $0.35\pm0.04$ \\
	 $\Sigma^{+} K^{+} \pi^{-}$       &  $4.44\pm0.52\pm0.25$ &$4.7\pm1.1\pm0.8$& $0.200\pm0.023\pm0.011 \pm0.011$   &    $0.21\pm0.06$               \\
	 $\Sigma^{+} K^{+} \pi^{-}\pi^{0}$  & $<2.5$ &                   -                   &$<$0.11      &-\\
	 $\Sigma^{+}  \phi$       &  $9.2\pm1.8\pm0.7$  &$8.5\pm1.2\pm1.2$& $0.414\pm0.080\pm0.030 \pm0.023$ & $0.39\pm0.06$\\
	 $\Sigma^{+} K^{+} K^{-}$(non-$\phi$) &$4.38\pm0.79\pm0.21$      &  -& $0.197\pm0.036\pm0.009 \pm0.011$ &-\\
	 
	 \hline\hline
	 \end{tabular}
	 \label{tab:vs}
	 \end{center}
	  \end{table}

\acknowledgments
\hspace{1.5em}
The BESIII collaboration thanks the staff of BEPCII and the IHEP computing center for their strong support. This work is supported in part by National Key Research and Development Program of China under Contracts Nos. 2020YFA0406400, 2020YFA0406300; Joint Large-Scale Scientific Facility Funds of the National Natural Science Foundation of China (NSFC) and Chinese Academy of Sciences (CAS) under Contracts Nos. U1932101, U1732263, U1832207; NSFC under Contracts Nos. 11625523, 11635010, 11675275, 11735014, 11822506, 11835012, 11935015, 11935016, 11935018, 11961141012, 11975021, 12022510, 12035009, 12035013, 12061131003, 12175321,  12221005; State Key Laboratory of Nuclear Physics and Technology, PKU under Grant No. NPT2020KFY04; the CAS Center for Excellence in Particle Physics (CCEPP); the CAS Large-Scale Scientific Facility Program; CAS Key Research Program of Frontier Sciences under Contract No. QYZDJ-SSW-SLH040; 100 Talents Program of CAS; Fundamental Research Funds for the Central Universities; INPAC and Shanghai Key Laboratory for Particle Physics and Cosmology; ERC under Contract No. 758462; European Union Horizon 2020 research and innovation programme under Contract No. Marie Sklodowska-Curie grant agreement No 894790; German Research Foundation DFG under Contracts Nos. 443159800, Collaborative Research Center CRC 1044, FOR 2359, FOR 2359, GRK 214; Istituto Nazionale di Fisica Nucleare, Italy; Ministry of Development of Turkey under Contract No. DPT2006K-120470; National Science and Technology fund; Olle Engkvist Foundation under Contract No. 200-0605; STFC (United Kingdom); The Knut and Alice Wallenberg Foundation (Sweden) under Contract No. 2016.0157; The Royal Society, UK under Contracts Nos. DH140054, DH160214; The Swedish Research Council; U. S. Department of Energy under Contracts Nos. DE-FG02-05ER41374, DE-SC-0012069.

\newpage
M.~Ablikim$^{1}$, M.~N.~Achasov$^{13,b}$, P.~Adlarson$^{73}$, R.~Aliberti$^{34}$, A.~Amoroso$^{72A,72C}$, M.~R.~An$^{38}$, Q.~An$^{69,56}$, Y.~Bai$^{55}$, O.~Bakina$^{35}$, I.~Balossino$^{29A}$
% ...
\\
\vspace{0.2cm}
(BESIII Collaboration)\\
\vspace{0.2cm} {\it
$^{1}$ Institute of High Energy Physics, Beijing 100049, People's Republic of China\\
% ...
$^{13}$ G.I. Budker Institute of Nuclear Physics SB RAS (BINP), Novosibirsk 630090, Russia\\
% ...
$^{29}$ INFN Sezione di Ferrara, (A)INFN Sezione di Ferrara, I-44122, Ferrara, Italy; (B)University of Ferrara, I-44122, Ferrara, Italy\\
% ...
$^{34}$ Johannes Gutenberg University of Mainz, Johann-Joachim-Becher-Weg 45, D-55099 Mainz, Germany\\
$^{35}$ Joint Institute for Nuclear Research, 141980 Dubna, Moscow region, Russia\\
% ...
$^{38}$ Liaoning Normal University, Dalian 116029, People's Republic of China\\
% ...
$^{55}$ Southeast University, Nanjing 211100, People's Republic of China\\
$^{56}$ State Key Laboratory of Particle Detection and Electronics, Beijing 100049, Hefei 230026, People's Republic of China\\
% ...
$^{69}$ University of Science and Technology of China, Hefei 230026, People's Republic of China\\
% ...
$^{72}$ University of Turin and INFN, (A)University of Turin, I-10125, Turin, Italy; (B)University of Eastern Piedmont, I-15121, Alessandria, Italy; (C)INFN, I-10125, Turin, Italy\\
$^{73}$ Uppsala University, Box 516, SE-75120 Uppsala, Sweden\\
% ...
$^{b}$ Also at the Novosibirsk State University, Novosibirsk, 630090, Russia\\
% ...

}


\begin{thebibliography}{99}
\bibitem{Klein:1989tj}
S.~R.~Klein,
Int. J. Mod. Phys. A \textbf{5}, 1457 (1990).

\bibitem{Asner:2008nq}
D.~M.~Asner \textit{et al.},
Int. J. Mod. Phys. A \textbf{24}, S1 (2009).

\bibitem{Yu:2017zst}
F.~S.~Yu, H.~Y.~Jiang, R.~H.~Li, C.~D.~L\"u, W.~Wang and Z.~X.~Zhao,
Chin. Phys. C \textbf{42}, 051001 (2018).

\bibitem{Rosner:2012gj}
J.~L.~Rosner,
Phys. Rev. D \textbf{86}, 014017 (2012).

\bibitem{Cheng:2015iom}
H.~Y.~Cheng,
Front. Phys. (Beijing) \textbf{10}, 101406 (2015).

\bibitem{BESIII:lmdenv}
M.~Ablikim \textit{et al.} (BESIII Collaboration), Phys.\ Rev.\ Lett.\ \textbf{115}, 221805 (2015).

\bibitem{BESIII:hadron}
M.~Ablikim \textit{et al.} (BESIII Collaboration), Phys.\ Rev.\ Lett.\  \textbf{116}, 052001 (2016).

\bibitem{BESIII:phh}
M.~Ablikim \textit{et al.} (BESIII Collaboration), Phys.\ Rev.\ Lett.\  \textbf{117}, 232002 (2016).






\bibitem{BESIII:lmdmunv}
M.~Ablikim \textit{et al.} (BESIII Collaboration), Phys.\ Lett.\ B \textbf{767}, 42 (2017).

\bibitem{BESIII:sigmapipi}
M.~Ablikim \textit{et al.} (BESIII Collaboration), Phys.\ Lett.\ B \textbf{772}, 388 (2017).

\bibitem{BESIII:nkspi}
M.~Ablikim \textit{et al.} (BESIII Collaboration), Phys. Rev. Lett. \textbf{118}, 112001 (2017).


\bibitem{BESIII:LmdX} 
M.~Ablikim \textit{et al.} (BESIII Collaboration), Phys.\ Rev.\ Lett. \textbf{121}, 062003 (2018).

\bibitem{BESIII:eX} 
M.~Ablikim \textit{et al.} (BESIII Collaboration), Phys.\ Rev.\ Lett. \textbf{121}, 251801 (2018).

\bibitem{BESIII:xik}
 M.~Ablikim \textit{et al.} (BESIII Collaboration), Phys.\ Lett.\ B \textbf{783}, 200 (2018).


\bibitem{BESIII:Lmdpieta} 
M.~Ablikim \textit{et al.} (BESIII Collaboration), Phys.\ Rev.\ D\ \textbf{99}, 032010 (2019).

\bibitem{BESIII:KsX} 
M.~Ablikim \textit{et al.} (BESIII Collaboration), Eur. Phys. J. C \textbf{80}, 935 (2020).

\bibitem{BESIII:pkseta}
 M.~Ablikim \textit{et al.} (BESIII Collaboration), Phys.\ Lett.\ B \textbf{817}, 136327 (2021).

\bibitem{BESIII:npi}
M.~Ablikim \textit{et al.} (BESIII Collaboration), Phys.\ Rev.\ Lett. \textbf{128}, 142001 (2022).


\bibitem{Belle:pkpi}
A.~Zupanc \textit{et al.} (Belle Collaboration), Phys. Rev. Lett. \textbf{113}, 042002 (2014).

\bibitem{Belle:pkpi_dcs}
S.~B.~Yang \textit{et al.} (Belle Collaboration), Phys. Rev. Lett. \textbf{117}, 011801 (2016).

\bibitem{Belle:sigmapipi} 
M.~Berger \textit{et al.} (Belle Collaboration), Phys.\ Rev.\ D \textbf{98}, 112006 (2018).



\bibitem{Belle:ppi0peta} 
S.~X.~Li \textit{et al.} (Belle Collaboration), Phys.\ Rev.\ D \textbf{103}, 072004 (2021).

\bibitem{Belle:pomega} 
S.~X.~Li \textit{et al.} (Belle Collaboration), Phys.\ Rev.\ D \textbf{104}, 072008 (2021).

\bibitem{Li:2021iwf}
H.~B.~Li and X.~R.~Lyu,
Natl. Sci. Rev. \textbf{8}, nwab181 (2021).

\bibitem{BESIII:2019odb}
M.~Ablikim \textit{et al.} (BESIII Collaboration),
Phys. Rev. D \textbf{100}, 072004 (2019).
\bibitem{Belle:2022uod}
L.~K.~Li \textit{et al.} (Belle Collaboration),
Sci. Bull. \textbf{68}, 583 (2023)

\bibitem{BESIII:2020kap}
M.~Ablikim \textit{et al.} (BESIII Collaboration),
Phys. Rev. D \textbf{103}, L091101 (2021).

\bibitem{LHCb:2019ldj}
R.~Aaij \textit{et al.} (LHCb Collaboration),
Phys. Rev. D \textbf{100}, 032001 (2019).

\bibitem{Belle-II:2022ggx}
F.~Abudin\'en \textit{et al.} (Belle-II Collaboration),
Phys. Rev. Lett. \textbf{130},  071802 (2023)



\bibitem{LHCb:pomega}
R.~Aaij \textit{et al.} (LHCb Collaboration), Phys. Rev. D \textbf{97}, 091101 (2018).



\bibitem{Cheng:2021qpd}
H.~Y.~Cheng,
Chin. J. Phys. \textbf{78}, 324 (2022).


\bibitem{Belle:2001hyr}
K.~Abe \textit{et al.} (Belle Collaboration),
Phys. Lett. B \textbf{524}, 33 (2002).

\bibitem{Cheng:1993gf}
H.~Y.~Cheng and B.~Tseng,
Phys. Rev. D \textbf{48}, 4188 (1993).

\bibitem{Guberina:2002fz}
B.~Guberina and H.~Stefancic,
Phys. Rev. D \textbf{65}, 114004 (2002).



\bibitem{Khoze:1983yp}
V.~A.~Khoze and M.~A.~Shifman,
Sov. Phys. Usp. \textbf{26}, 387 (1983).



\bibitem{LHCb:2018nfa}
R.~Aaij \textit{et al.} (LHCb Collaboration),
Phys. Rev. Lett. \textbf{121}, 092003 (2018).





\bibitem{LHCb:2021vll}
R.~Aaij \textit{et al.} (LHCb Collaboration),
Sci. Bull. \textbf{67}, 479 (2022).

\bibitem{Cheng:2021vca}
H.~Y.~Cheng,
Sci. Bull. \textbf{67}, 445 (2022).




\bibitem{Cen:2019ims}
J.~Y.~Cen, C.~Q.~Geng, C.~W.~Liu and T.~H.~Tsai,
Eur. Phys. J. C \textbf{79}, no.11, 946 (2019).

\bibitem{Hsiao:2019yur}
Y.~K.~Hsiao, Y.~Yao and H.~J.~Zhao,
Phys. Lett. B \textbf{792}, 35-39 (2019). 

\bibitem{CLEO:1993fhs}
P.~Avery \textit{et al.} (CLEO Collaboration),
Phys. Rev. Lett. \textbf{71}, 2391 (1993).

\bibitem{BESIII:2022ulv}
M.~Ablikim \textit{et al.} (BESIII Collaboration),
Chin. Phys. C \textbf{46}, 113003 (2022).

\bibitem{Ablikim:2009aa}
  M.~Ablikim {\it et al.} (BESIII Collaboration),
  Nucl.\ Instrum.\ Meth.\ A {\bf 614}, 345 (2010).

\bibitem{Yu:IPAC2016-TUYA01}
   C.~H.~Yu {\it et al.},
  Proceedings of IPAC2016, Busan, Korea, (2016).
  
  
  \bibitem{Ablikim:2019hff}
  M.~Ablikim {\it et al.} (BESIII Collaboration),
  Chin. Phys. C {\bf 44}, 040001 (2020).

\bibitem{etof}
 X.~Li {\it et al.}, 
 Radiat. Detect. Technol. Methods {\bf 1}, 13 (2017).
 \bibitem{etof2}
 Y.~X.~Guo {\it et al.},
 Radiat. Detect. Technol. Methods {\bf 1}, 15 (2017).
 \bibitem{etof3}
 P.~Cao {\it et al.}, Nucl.\ Instrum.\ Meth.\ A {\bf 953}, 163053 (2020).

\bibitem{geant4}
  S.~Agostinelli {\it et al.} (GEANT4 Collaboration),
  Nucl.\ Instrum.\ Meth.\ A {\bf 506}, 250 (2003).

 \bibitem{bes:boost}
Z.~Y.~Deng {\it et al.},
Chin.\ Phys.\ C  {\bf 30}, 371 (2006).  

\bibitem{geo2}
Z.~Y.~You, Y.~T.~Liang and Y.~J.~Mao, 
Chin.\ Phys.\ C  {\bf 32}, 572 (2008).

\bibitem{detvis}
  K.~X.~Huang, {\it et al.},
  Nucl.\ Sci.\ Tech. {\bf 33}, 142 (2022).

\bibitem{ref:kkmc}
  S.~Jadach, B.~F.~L.~Ward and Z.~Was,
  Phys.\ Rev.\ D {\bf 63}, 113009 (2001).

\bibitem{ref:kkmc2}
S.~Jadach, B.~F.~L.~Ward and Z.~Was,
Comput. Phys. Commun. \textbf{130}, 260 (2000).


\bibitem{ref:evtgen}
D.~J.~Lange,
Nucl. Instrum. Meth. A \textbf{462}, 152 (2001).


\bibitem{pdg:2020}
R.~L.~Workman {\it et al.} (Particle Data Group),
Prog.\ Theor.\ Exp.\ Phys. \textbf{2022}, 083C01 (2022).

\bibitem{ref:lundcharm}
  J.~C.~Chen {\it et al.},
  Phys.\ Rev.\ D {\bf 62}, 034003 (2000).

\bibitem{ref:lundcharm2}
R.~L.~Yang, R.~G.~Ping and H.~Chen,
  Chin.\ Phys.\ Lett.\  {\bf 31}, 061301 (2014).

\bibitem{photos}
E.~Richter-Was,
Phys. Lett. B \textbf{303}, 163 (1993).

\bibitem{ARGUS:1990hfq}
H.~Albrecht \textit{et al.} (ARGUS Collaboration),
Phys. Lett. B \textbf{241}, 278 (1990).

\bibitem{PWA}
Y.~Jiang {\it et al.},
frameworkTF-PWApackage, GitHublink:https://github.com/jiangyi15/tf-pwa (2020).
  

\end{thebibliography}
\end{document}